# Quantum transport simulation of exciton condensate transport physics in a double layer graphene system

Xuehao Mou,[1] Leonard F. Register,[1] Allan H. MacDonald,[2] and Sanjay K. Banerjee[1]

[1]*Department of Electrical and Computer Engineering and Microelectronics Research Center, The University of Texas at Austin, Austin, Texas 78758, United States*

[2]*Department of Physics, The University of Texas at Austin, Austin, Texas 78712, United States*

*Abstract*—Spatially indirect electron-hole exciton condensates stabilized by interlayer Fock exchange interactions have been predicted in systems containing a pair of two-dimensional semiconductor or semimetal layers separated by a thin tunnel dielectric. The layer degree of freedom in these systems can be described as a *pseudospin*. Condensation is then analogous to ferromagnetism, and the interplay between collective and quasiparticle contributions to transport is analogous to phenomena that are heavily studied in spintronics. These phenomena are the basis for *pseudospintronic* device proposals based on possible low-voltage switching between high (nearly shorted) and low interlayer conductance states and on near perfect Coulomb drag-counterflow current along the layers. In this work, a quantum transport simulator incorporating a non-local Fock exchange interaction is presented, and used to model the essential transport physics in the bilayer graphene system. Finite size effects, Coulomb drag-counterflow current, critical interlayer currents beyond which interlayer DC conductance collapses at sub-thermal voltages, non-local coupling between interlayer critical currents in multiple lead devices, and an Andreev-like reflection process are illustrated.

## I. INTRODUCTION

Coherent pairing of electrons and holes localized in separate III-V semiconductor layers has been observed at high magnetic fields and cryogenic temperatures in systems containing a pair of quantum wells separated by tunnel barriers (where holes are defined within as empty states in partially empty Landau levels).[1-4] In these spatially indirect exciton condensates, interlayer coherence is mediated by the interlayer Fock exchange interactions.[3] The consequences of condensate formation that have been studied experimentally include novel transport effects such as near perfect Coulomb drag-counterflow currents, and greatly enhanced interlayer conductance up to a critical current $I_c$ and a corresponding critical interlayer voltage $V_c$ beyond which the DC current collapses,[5] the latter effect being partially analogous to transport phenomena in Josephson junctions. In this paper, we present a theoretical quantum transport simulation study of closely related phenomena in the absence of a magnetic field. Although spatially indirect exciton condensation has not yet been observed in this regime, it is expected to appear when nesting is established between electron- and hole-like Fermi surfaces of two-dimensional semiconductors, gapless semiconductors, or semimetals. The prospects for observing spatially indirect exciton condensation are therefore being enhanced by progress in fabricating and studying two-dimensional materials. In this study we choose for the sake of definiteness to study graphene bilayers, but our conclusions apply equally well to other systems.

First successfully isolated about a decade ago, graphene already has exhibited a broad set of novel phenomena of interest to researchers from multiple disciplines, such as the half-integer quantum Hall effect and the related Berry's phase that had been predicted theoretically by physicists.[6,7] Its unique symmetric and conical band structure has allowed research on the properties of massless Dirac Fermions, including novel transport properties in semiconductor device physics. When separated by dielectric tunnel barriers such has hexagonal boron nitride, it is possible to take advantage of the symmetry between graphene's electron and hole band structures to establish the Fermi surface nesting conditions that favor interlayer exciton condensates.[8,9] Systems based on monolayer or few-layer transition metal dichalcogenide (TMD) semiconductors have extremely large exciton binding energies and may provide an even better platform for such condensates. The prospect of transport effects analogous to those in already studied in III-V quantum well systems, but absent magnetic fields and at higher temperatures[9,10] perhaps even room temperature, have led to novel "beyond CMOS" device proposals with switching energies potentially of only a few tens of zepto-joules, including the Bilayer pseudoSpin Field-Effect Transistor (BiSFET)[11-14] and, more recently, the Bilayer pseudoSpin Junction Transistor (BiSJT).[15]

Studies of the transport properties of systems containing a spatially indirect exciton condensate require simulators that incorporate the non-local Fock exchange interaction. We describe such a simulator in detail, and use it to model essential transport physics including finite size effects, critical currents at sub-thermal voltages, Coulomb drag-counterflow current, non-local coupling between critical currents in multi-terminal devices, and a reflection process for this system akin to the Andreev reflection at the boundary of a conventional superfluid. Although the simulation techniques described here should be transferable to TMD based system, or to other two-dimensional semiconductors or semimetals, simulations might in some cases have significantly greater computational costs, depending on the complexity of the electronic structure. The purpose of this work is to model essential transport properties *in the presence of* the exciton condensates, including but not limited to those serving as the basis for BiSFET and BiSJT. The challenges which must be met to achieve such condensates have been and are being addressed elsewhere.[8-10,16] This work is intended to motivate such work, and to help with the interpretation and design of experiments as well as potential devices.

Portions of this work were reported previously in conference publications.[13,17,18] Here, more details of the method, more complete sets of data, a more comprehensive and synergistic analysis thereof, and a more thorough discussion of the underlying essential physics including interpretations in terms of electrons, holes, and excitons are provided.

The quantum transport simulator and the system model used here are discussed in Section II; various transport properties and discussions of underlying essential physics are addressed in Section III.

## II. TIGHT-BINDING HARTREE-FOCK MODEL

Although we discuss electron-hole interactions, transport simulations are performed entirely in terms of electron states within the conduction and valence bands of opposite layers. The Hartree-Fock approximation is used to model the exchange interaction between electrons (due to their indistinguishability and odd parity) within a single-particle framework. The resulting time-independent Schrödinger's equation for single-electron energy eigenstates $\{\varphi_\beta(\mathbf{r}), E_\beta\}$, neglecting for the moment the external potential profile, takes the form:

$$-\frac{\hbar^2}{2m}\nabla^2\varphi_\beta(\mathbf{r})+\varphi_\beta(\mathbf{r})\int d\mathbf{r}'V_C(\mathbf{r},\mathbf{r}')\rho(\mathbf{r}',\mathbf{r}')-\int d\mathbf{r}'V_C(\mathbf{r},\mathbf{r}')\rho(\mathbf{r}',\mathbf{r})\varphi_\beta(\mathbf{r}')=E_\beta\varphi_\beta(\mathbf{r}), \qquad (1a)$$

where $\rho(\mathbf{r},\mathbf{r}')$ is the density matrix,

$$\rho(\mathbf{r},\mathbf{r}')\equiv\sum_{\beta'}f_{\beta'}\varphi_{\beta'}(\mathbf{r})\varphi_{\beta'}^*(\mathbf{r}'). \qquad (1b)$$

The second term on the left of Eq. (1a) is the local Hartree potential term describing the classical electrostatic potential at $\mathbf{r}$ due to a charge distribution $-q\rho(\mathbf{r}) = -q\rho(\mathbf{r},\mathbf{r})$, to which we can readily add the contributions of externally applied electrostatic fields. The third term is the non-local and purely quantum mechanical Fock exchange interaction, which lowers the electron-electron interaction energy due to tendency of the odd-parity electrons to be further away from each other than expected classically.[19] The indices $\beta$ ($\beta'$) label the single electron states and the $f_\beta$ ($f_{\beta'}$) are their occupation probabilities. $V_C(\mathbf{r},\mathbf{r}')$ is the Coulomb interaction energy between electrons at positions $\mathbf{r}$ and $\mathbf{r}'$. (Notably, the contributions to the Hartree and Fock terms for $\beta = \beta'$ cancel, demonstrating that Eq. (1) avoids interaction between any electron and itself.)

For transport simulations, we switch to an atomistic tight-binding approximation considering in this work just one $2p_z$ ($\pi$) orbital per carbon atom site located at discrete positions $\mathbf{R}$. (More general models are also described by the equations that follow if the indices $\mathbf{R}$ are considered to also represent the various orbitals on each atom.) Within this tight-binding framework, Eq. (1) can be written in the following form[20]:

$$H_{TB}\varphi_\beta(\mathbf{R}_T)+V_H(\mathbf{R}_T)\varphi_\beta(\mathbf{R}_T)+\sum_{\mathbf{R}_B}V_F(\mathbf{R}_T,\mathbf{R}_B)\varphi_\beta(\mathbf{R}_B)=E_\beta\varphi_\beta(\mathbf{R}_T), \qquad (2a)$$

$$V_F(\mathbf{R}_T,\mathbf{R}_B)\equiv -V_C(\mathbf{R}_T,\mathbf{R}_B)\rho(\mathbf{R}_T,\mathbf{R}_B), \qquad (2b)$$

$$\rho(\mathbf{R}_T,\mathbf{R}_B)\equiv\sum_{\beta'}f_{\beta'}\varphi_{\beta'}(\mathbf{R}_T)\varphi_{\beta'}^*(\mathbf{R}_B), \qquad (2c)$$

for the top (T) layer. The form for the bottom (B) layer is obtained by switching the "T" and "B" indices. We have explicitly considered only the *inter*layer Fock exchange interaction, labeled $V_F$, which is the basis for exciton condensation. $V_H$ is the nominal Hartree potential energy, which in principle incorporates contributions due to charges on either layer, any external charge (externally applied fields), and, here, any uniform correction to the energy due to *intra*-layer Fock exchange interactions. $H_{TB}$ includes the *intra*-layer and (much weaker) *inter*layer "bare" tight-binding hopping energies. Note that $V_F(\mathbf{R}_T,\mathbf{R}_B)$ can become complex under non-equilibrium conditions through the density matrix $\rho(\mathbf{R}_T,\mathbf{R}_B)$.

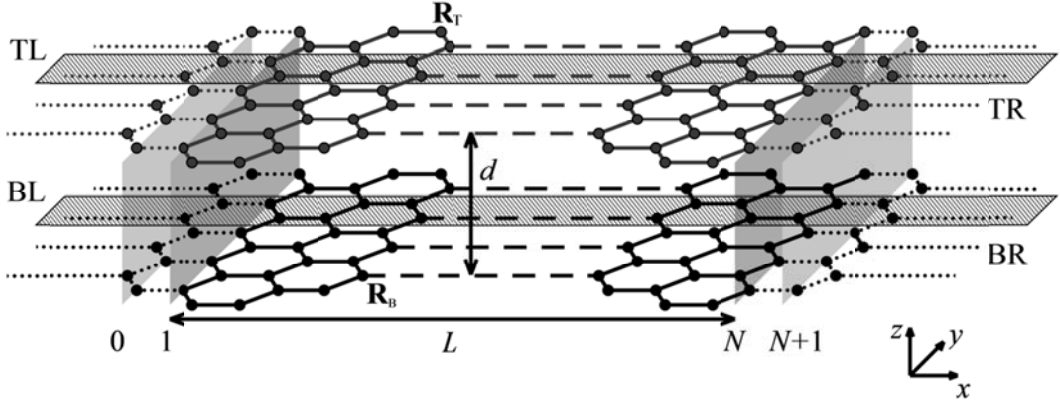

FIG. 1. Simulated structure. Two graphene layers (infinite in the $y$ direction) are coupled via bare and Fock exchange interactions in a channel of length $L$ and with interlayer separation $d$ across a tunnel dielectric. The effective dielectric constant $\varepsilon_r$ represents the dielectric environment between and, still more importantly, above and below the graphene layers as a whole. The periodicity in the $y$ direction is represented by the shaded stripes along the transport direction encompassing one armchair chain of graphene atom per layer. Atoms within the armchair chain are labeled by consecutive integers, starting from 1 on the left of the channel and ending with $N$ on the right. Two consecutive atomic sites within one armchair chain of atoms and within the same layer correspond to a primitive unit cell of monolayer graphene. Four semi-infinite, perfect absorbing/injecting leads are modeled, and labeled TL (top left), BL (bottom left), TR (top right) and BR (bottom right), which conceptually (they do appear only implicitly in the boundary conditions) incorporate atoms 0, −1, −2, … on the left and $N+1$, $N+2$, $N+3$, … on the right within a layer.

Fig. 1 illustrates the four-terminal structure used in the simulator developed in Ref. 17, where two graphene sheets are separated by a distance $d$ through an interlayer/tunnel dielectric and coupled via the quasi-single-particle bare coupling and the many-body Fock exchange interaction, both within a channel of length $L$. The dielectric properties of the environment, including the interlayer dielectric, the more important dielectrics above and below the condensate region, and free-carrier screening, are all represented very roughly by a single effective dielectric constant $\varepsilon_r$, given our focus on essential transport physics in the presence of the exciton condensate and not on the conditions required to create the condensate. Beyond the channel, the interlayer Fock exchange interaction is taken to be zero, qualitatively consistent with a designed increase in the interlayer separation or in the permittivity of the dielectric environment to localize the condensate. The width of the graphene sheets perpendicular to the transport direction are taken to be infinite for simulation purposes. For specificity, the graphene is oriented such that there is an armchair pattern of atoms along the transport ($x$) direction. The pattern of bare interlayer tunnel-coupling through the dielectric could be quite complex in principle (we do not specify a particular dielectric here), and will not in general correspond to the "A-B" coupling pattern between A and B sublattice atoms of opposite graphene layers as in a Bernal stack. Moreover, given a mixed bare interlayer coupling pattern, bulk calculations have shown that the coherent interlayer exciton condensate would most readily couple to the A-A (or, equivalently, B-B) component of the bare interlayer coupling pattern.[21] Therefore, we have taken the bare interlayer coupling to be of the latter A-A form. The carriers in both layers are conceptually created via external gates that are omitted in Fig. 1. For simulation purposes, the required carrier concentrations are adjusted by applying fixed electrostatic potential energies $V(\mathbf{R}_T) = V_T = -V(\mathbf{R}_B) = -V_B = -V_{\text{diff}}/2$, with the energy reference being the equilibrium (no voltages applied) chemical potential (Fermi level). Consequently, the nominal electron and hole concentrations are equal given the symmetric band structure of graphene. For all simulations, the variations in carrier concentration with applied voltages are negligible compared to the nominal concentrations, and hence only the Fock interaction, $V_F(\mathbf{R}_T,\mathbf{R}_B)$, is considered self-consistently. For this transport system, the single-electron state label $\beta = \{\gamma,\sigma,k_{x,i},k_y,s\}$ identifies the lead of injection $\gamma$, the energy "valley" $\sigma$, the incident crystal momentum $\hbar k_{x,i}$ of the injected wave, the transverse mode of well-defined crystal momentum $\hbar k_y$, and (real) spin $s$. The occupation probability $f_\beta$ is determined by its energy and the voltage applied to the lead in which it originates, $V_\gamma = -\mu_\gamma/q$, where $\mu_\gamma$ is the chemical potential (Fermi level) of the lead $\gamma$.

As shown by the stripes in the top and bottom layer oriented parallel to the transport direction in Fig. 1, the simulated structure is periodic in the transverse ($y$) direction with a lattice constant of $a_y$ encompassing a single armchair pattern of carbon atoms with atomic locations within each layer $\mathbf{\Gamma}_{T(B)}$, a subset of $\mathbf{R}_{T(B)}$. (Note while we assumed A-A coupling here, in general there would be no requirement that the atoms on the top and bottom layers be vertically aligned for this purpose.) Therefore, the Fock exchange interaction term can be rewritten for, e.g., any $\mathbf{\Gamma}_T$ as

$$\begin{aligned}
\sum_{\mathbf{R}_B} V_F(\mathbf{\Gamma}_T,\mathbf{R}_B)\varphi_\beta(\mathbf{R}_B) &= -\sum_{\mathbf{\Gamma}_B}\sum_{\eta=-\infty}^{\infty} V_C(\mathbf{\Gamma}_T,\mathbf{\Gamma}_B+\eta a_y\hat{\mathbf{y}})\rho(\mathbf{\Gamma}_T,\mathbf{\Gamma}_B+\eta a_y\hat{\mathbf{y}})\varphi_\beta(\mathbf{\Gamma}_B+\eta a_y\hat{\mathbf{y}}) \\
&= -\sum_{\mathbf{\Gamma}_B}\sum_{\eta=-\infty}^{\infty} V_C(\mathbf{\Gamma}_T,\mathbf{\Gamma}_B+\eta a_y\hat{\mathbf{y}})\sum_{\beta'} f_{\beta'}\varphi_{\beta'}(\mathbf{\Gamma}_T)\varphi_{\beta'}^*(\mathbf{\Gamma}_B)e^{-ik_y'\eta a_y}\varphi_\beta(\mathbf{\Gamma}_B)e^{ik_y\eta a_y} \\
&= \sum_{\mathbf{\Gamma}_B}\sum_{\beta'} V_{C,k_y-k_y'}(\mathbf{\Gamma}_T,\mathbf{\Gamma}_B) f_{\beta'}\varphi_{\beta'}(\mathbf{\Gamma}_T)\varphi_{\beta'}^*(\mathbf{\Gamma}_B)\varphi_\beta(\mathbf{\Gamma}_B) \\
&= \sum_{\mathbf{\Gamma}_B} V_{F,k_y}(\mathbf{\Gamma}_T,\mathbf{\Gamma}_B)\varphi_\beta(\mathbf{\Gamma}_B)
\end{aligned} \quad (3a)$$

where

$$V_{C,k_y-k_y'}(\mathbf{\Gamma}_T,\mathbf{\Gamma}_B) \equiv \sum_{\eta=-\infty}^{\infty} -V_C(\mathbf{\Gamma}_T,\mathbf{\Gamma}_B+\eta a_y\hat{\mathbf{y}})e^{i(k_y-k_y')\eta a_y} \quad (3b)$$

and

$$V_{F,k_y}(\mathbf{\Gamma}_T,\mathbf{\Gamma}_B) \equiv \sum_{\beta'} V_{C,k_y-k_y'}(\mathbf{\Gamma}_T,\mathbf{\Gamma}_B) f_{\beta'}\varphi_{\beta'}(\mathbf{\Gamma}_T)\varphi_{\beta'}^*(\mathbf{\Gamma}_B). \quad (3c)$$

Here, $\hat{\mathbf{y}}$ is a unit vector in the $y$ direction and $\eta$ is an integer. Moreover, absent explicit free-carrier screening as modeled here, the Coulomb interaction potential energy is just

$$V_C(\mathbf{\Gamma}_T,\mathbf{\Gamma}_B+\eta a_y\hat{\mathbf{y}}) = \frac{q^2}{4\pi\varepsilon_r|\mathbf{\Gamma}_T-(\mathbf{\Gamma}_B+\eta a_y\hat{\mathbf{y}})|}, \quad (4)$$

which allows the $V_{C,k_y-k_y'}(\mathbf{\Gamma}_T,\mathbf{\Gamma}_B)$ to be pre-calculated and stored. Using Eq. (3), Eq. (2) can be rewritten in the quasi-one-dimensional (quasi-1D) form

$$H_{TB}\varphi_\beta(\mathbf{\Gamma}_T)+V_H(\mathbf{\Gamma}_T)\varphi_\beta(\mathbf{\Gamma}_T)+\sum_{\mathbf{\Gamma}_B} V_{F,k_y}(\mathbf{\Gamma}_T,\mathbf{\Gamma}_B)\varphi_\beta(\mathbf{\Gamma}_B) = E_\beta\varphi_\beta(\mathbf{\Gamma}_T) \quad (5)$$

for the top layer. A similar expression can be obtained for the bottom layer.

*For each $\beta$*—suppressing the indices for notational convenience now—Eq. (5) can be written in matrix form including both top and bottom layers as

$$E\boldsymbol{\varphi}_m - \sum_n \mathbf{H}_{m,n}\boldsymbol{\varphi}_n = 0. \tag{6a}$$

Here, $m(n)$ label individual atomic slices along the transport direction, i.e., pairs of atoms, one in the top layer and one in the bottom, along the shaded armchair atomic chains in the transport direction in Fig. 1. The $\boldsymbol{\varphi}_{m(n)}$ are the corresponding tight-binding 2×1 matrices/column vectors for each slice $m(n)$ with components $\varphi_{T,m(n)}$ and $\varphi_{B,m(n)}$. (Here we have assumed that there is nothing to induce coupling between orthogonal spin states in this graphene-based system, so that we may treat each spin state separately, unlike what would be the case for TMDs.) The $\mathbf{H}_{m,n}$ are 2×2 tight-binding potential matrices coupling slices $m$ and $n$ via both the bare and Fock interactions of Eq. (5). Note that there are only non-zero onsite and nearest-neighbor intra-layer and interlayer bare coupling interactions, while the interlayer Fock interactions extend among all sites within the channel but do not, in this model, couple to points beyond the channel.

The boundary conditions at the ends of the channel are obtained by assuming that the four leads, TL, BL, TR and BR, are semi-infinite and perfectly absorbing. Labeling the atomic sites *within* the channel from $m = 1$ to $N$, the incident (i), if any, and "outgoing" (o) (including outwardly evanescing or reflected into the lead of incidence) components of the wavefunctions are related across the simulation region boundaries—between slices 0 and 1 on the left boundary for example—by

$$\boldsymbol{\varphi}_0 = \boldsymbol{\varphi}_{0;i} + \boldsymbol{\varphi}_{0;o} = \boldsymbol{\varphi}_{0;i} + \mathbf{P}_{1,0;o}\boldsymbol{\varphi}_{1;o} = \boldsymbol{\varphi}_{0;i} + \mathbf{P}_{1,0;o}(\boldsymbol{\varphi}_1 - \boldsymbol{\varphi}_{1;i}) = \mathbf{P}_{1,0;o}\boldsymbol{\varphi}_1 + (\mathbf{I} - \mathbf{P}_{1,0;o}\mathbf{P}_{0,1;i})\boldsymbol{\varphi}_{0;i}. \tag{7}$$

Here, $\mathbf{P}_{0,1;i}$ and $\mathbf{P}_{1,0;o}$ are 2×2 diagonal matrices that relate the complex amplitudes of the incident (i) and outgoing (o) wave-function components between adjacent slices. The $\mathbf{P}_{0,1;i}$ and $\mathbf{P}_{1,0;o}$ are defined consistent with the required Bloch function form of the propagating modes in the leads, or the counterparts of Bloch functions with complex wave-vectors for "outgoing" evanescent modes *into* the leads. For example, for incident (outgoing) propagating wave-functions $\varphi_{\gamma'',m,i(o)}$ in a given lead $\gamma''$ (with $m \leq 1$ or $m \geq N$ for the left and right leads, respectively), $\varphi_{\gamma'',m,i(o)} = e^{i\mathbf{k}_{\gamma'',i(o)} \cdot \mathbf{\Gamma}_{\gamma'',m}} u_{\mathbf{k}_{\gamma'',i(o)}}(\mathbf{\Gamma}_{\gamma'',m})$, where $u_{\mathbf{k}_{\gamma'',i(o)}}(\mathbf{\Gamma}_{\gamma'',m})$ has the periodicity of the unit cell (four slices in the transport direction). The $\mathbf{k}_{\gamma'',i(o)}$ are readily determined based on the energy $E$, potential energy in the given lead $\gamma''$, transverse wave-vector $k_y$ and required direction of propagation. Applying Eq. (7) to the Schrodinger's equation, Eq. (6a), yields for the left boundary (slices 0 and 1),

$$E\mathbf{I}\boldsymbol{\varphi}_1 - (\mathbf{H}_{1,2}\boldsymbol{\varphi}_2 + \mathbf{H}_{1,1}\boldsymbol{\varphi}_1) - \mathbf{H}_{1,0}\mathbf{P}_{1,0;o}\boldsymbol{\varphi}_1 = \mathbf{H}_{1,0}(\mathbf{I} - \mathbf{P}_{1,0;o}\mathbf{P}_{0,1;i})\boldsymbol{\varphi}_{0;i}. \tag{6b}$$

A similar expression is obtained for the right boundary (slices $N$ and $N+1$),

$$E\mathbf{I}\boldsymbol{\varphi}_N - (\mathbf{H}_{N,N-1}\boldsymbol{\varphi}_{N-1} + \mathbf{H}_{N,N}\boldsymbol{\varphi}_N) - \mathbf{H}_{N,N+1}\mathbf{P}_{N,N+1;o}\boldsymbol{\varphi}_N = \mathbf{H}_{N,N+1}(\mathbf{I} - \mathbf{P}_{N,N+1;o}\mathbf{P}_{N+1,N;i})\boldsymbol{\varphi}_{N+1;i}. \tag{6c}$$

The injected wave-functions, $\boldsymbol{\varphi}_{0;i}$ and $\boldsymbol{\varphi}_{N+1;i}$, are chosen to be localized to either the left end or the right end of the channel and to the top or bottom layer consistent with the definition of the index $\gamma$, and are normalized to carry the appropriate amount of incident current per transverse mode $k_y$ per unit energy consistent with Landauer-Büttiker theory.[22] Starting with an initial guess of the modified Hartree-Fock potentials $V_{F,k_y}(\Gamma_T,\Gamma_B)$, if just uniformly zero, we solve Eqs. (6a-c) for each value of $k_y$ independently. The Fock exchange interaction for each value of $k_y$ then must be recalculated as a function of *all* values of $k_y$. We repeat this process iteratively to obtain a self-consistent solution (analogous to self-consistent Schrödinger-Poisson's calculations, except with non-local potential or off-diagonal density matrix elements for the exchange interactions).

For reference, Eqs. (6a-c) are equivalent to the NEGF problem $\boldsymbol{\varphi} = \mathbf{GS}$, where the Green's function is $\mathbf{G} = (E\mathbf{I} - \mathbf{H} - \boldsymbol{\Sigma})^{-1}$ with self-energy matrix $\boldsymbol{\Sigma}$ and source vector $\mathbf{S}$.[22] Therefore, $\boldsymbol{\varphi} = \mathbf{GS}$ can be rewritten as $(E\mathbf{I} - \mathbf{H} - \boldsymbol{\Sigma})\boldsymbol{\varphi} = \mathbf{S}$. Comparing this latter form to Eqs. (6a-c) allows us to identify $\boldsymbol{\Sigma}_{1,1} = \mathbf{H}_{1,0}\mathbf{P}_{1,0;o}$, $\boldsymbol{\Sigma}_{N,N} = \mathbf{H}_{N,N+1}\mathbf{P}_{N,(N+1);o}$, and all other $\boldsymbol{\Sigma}_{m,n} = \mathbf{0}$. Similarly, $\mathbf{S}_1 = \mathbf{H}_{1,0}(\mathbf{I} - \mathbf{P}_{1,0;o}\mathbf{P}_{0,1;i})\boldsymbol{\varphi}_{0;i}$, $\mathbf{S}_N = \mathbf{H}_{N,N+1}(\mathbf{I} - \mathbf{P}_{N,N+1;o}\mathbf{P}_{N+1,N;i})\boldsymbol{\varphi}_{N+1;i}$, and all other $\mathbf{S}_l = \mathbf{0}$. However, in these ballistic transport equations, we have no need for the full $\mathbf{G}$ matrix, which includes terms representing transport between internal points.

In the tight-binding model, the charge current flow from a top layer atomic site $\mathbf{R}_T$ to a bottom layer site $\mathbf{R}_B$ associated with any particular state $\beta$, $I_\beta(\mathbf{R}_T,\mathbf{R}_B)$, is

$$I_\beta(\mathbf{R}_T,\mathbf{R}_B) = (-2q/h)\mathrm{Im}\left\{-f_\beta \varphi_\beta^*(\mathbf{R}_T)[H_b(\mathbf{R}_T,\mathbf{R}_B) + V_F(\mathbf{R}_T,\mathbf{R}_B)]\varphi_\beta(\mathbf{R}_B)\right\}, \quad (8)$$

where the coupling between sites $\mathbf{R}_B$ and $\mathbf{R}_T$ is decomposed in terms of the bare coupling $H_b$ and the Fock exchange interaction $V_F$. Spin-degeneracy is included with non-interacting spins. The total site-to-site current $I(\mathbf{R}_T,\mathbf{R}_B)$, therefore can be written as

$$\begin{aligned}I(\mathbf{R}_T,\mathbf{R}_B) &= \sum_\beta I_\beta(\mathbf{R}_T,\mathbf{R}_B) \\ &= (-2q/h)\mathrm{Im}\left\{-\sum_\beta f_\beta \varphi_\beta^*(\mathbf{R}_T)[H_b(\mathbf{R}_B,\mathbf{R}_T) + V_F(\mathbf{R}_B,\mathbf{R}_T)]\varphi_\beta(\mathbf{R}_B)\right\}\end{aligned}. \quad (9)$$

Upon inspection, it can be seen that for the total current, but only for the total current, the component associated with the Fock exchange interaction includes the imaginary part of the product of the interlayer density matrix (Eqs. (2b-c)) with its Hermitian adjoint, which intrinsically vanishes. The expression for the *total* interlayer site-to-site current $I(\mathbf{R}_T,\mathbf{R}_B)$ thus takes the simplified form,

$$\begin{aligned}I(\mathbf{R}_T,\mathbf{R}_B) &= (-2q/h)H_b(\mathbf{R}_T,\mathbf{R}_B)\mathrm{Im}\left\{\sum_\beta f_\beta \varphi_\beta(\mathbf{R}_T)\varphi_\beta^*(\mathbf{R}_B)\right\}, \\ &= (-2q/h)H_b(\mathbf{R}_T,\mathbf{R}_B)|\rho(\mathbf{R}_T,\mathbf{R}_B)|\sin\{\arg[\rho(\mathbf{R}_T,\mathbf{R}_B)]\}\end{aligned} \quad (10)$$

a result that is independent of condensate formation.[14,23] Therefore, bare coupling between sites is required for a nonzero *net* current flow between those sites, and, therefore, a so-called "spontaneous" condensate formed in the absence of any bare interlayer coupling—with a large interlayer density matrix $\rho(\mathbf{R}_T,\mathbf{R}_B)$ self-consistently obtained via Fock interlayer exchange interaction that is proportional to $\rho(\mathbf{R}_T,\mathbf{R}_B)$—is incapable of carrying a net interlayer current. However, the enhanced interlayer density matrix $\rho(\mathbf{R}_T,\mathbf{R}_B)$ in the presence of a condensate can enhance the interlayer current flow. Note that, for a particular state $\beta$, non-zero current $I_\beta(\mathbf{R}_T,\mathbf{R}_B)$ still can be carried between sites with no associated bare coupling.

## III. SIMULATION RESULTS

In this section, we begin showing the formation of the condensate in nanoscale structures, and with calculations of intra-layer and interlayer transmission probabilities under equilibrium conditions, $V_{TL} = V_{BL} = V_{TR} = V_{BR} = 0$ V, in the presence of the condensate, and for reference also in its absence. However, as will become apparent, simply knowing these transmission probabilities is not enough to predict all essential features of the non-equilibrium transport. We then simulate non-equilibrium transport under three biasing conditions: a "drag-counterflow" biasing condition with only one layer biased, $V_{BL} = -V_{BR} = V_{al}/2$ and $V_{TL} = V_{TR} = 0$, analogous to that previously explored in Ref. 24 within a simpler 1-D effective mass model with a local exchange interaction; an "interlayer" biasing condition much like that which was the basis for previously proposed BiSFET with only one end biased,[11,12] $V_{TL} = -V_{BL} = V_{il}/2$ and $V_{TR} = V_{BR} = 0$; and a "mixed" biasing scheme that has become the basis for the separately proposed BiSJT,[15] $V_{TL} = -V_{BL} = V_{ctrl}/2$ and $V_{TR} = -V_{BR} = V_{drv}/2$, also touched upon in Ref. 24.

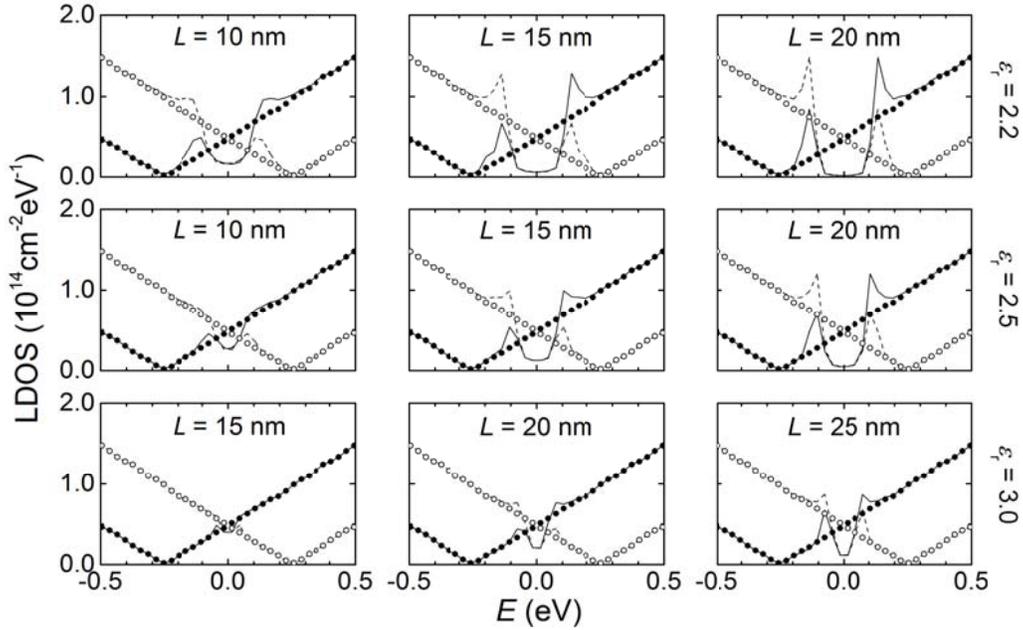

FIG. 2. Local density of state (LDOS) plotted in the center of the channel. The top-layer LDOS is plotted with solid lines and bottom-layer LDOS with dashed lines, respectively with solid and open circles representing the condensate-free LDOS for comparison. A band gap starts to form and saturates to its bulk value with increasing channel length. For stronger condensates (low $\varepsilon_r$), the (incomplete) band gap saturates earlier, and is larger for the same channel length.

### A. Formation of a nanoscale condensate

In this subsection, we illustrate how the condensate forms in nanoscale channels of increasing length. (However, scaling issues *per se* will be addressed subsequently.) The calculated local density of states (LDOS) in the center of the channel for the structure in Fig. 1 under equilibrium conditions is plotted in Fig. 2 for various channel lengths $L$ and effective dielectric constants $\varepsilon_r$, with carrier concentrations $n$ (top layer) = $p$ (bottom layer) ≈ $6 \times 10^{12}$ cm$^{-2}$ ($V_{\text{diff}}$ = 0.5 eV) and interlayer separation $d = 1$ nm. For these simulations—and all others to follow in this work—room temperature (300 K) was assumed in calculating the thermal distributions $f_\beta$ of injected carriers from the leads. For increasing channel lengths, a band gap, if incomplete, begins to form about the chemical potential $\mu = 0$ eV, which is also the point at which the top and bottom layer Dirac cones would otherwise cross, as indicated by the reduction in the LDOS. The edges of the band gap region saturate toward bulk results[20,21] (although less obviously so for $\varepsilon_r$ = 3.0 and the channel lengths considered here). The degree to which the gap forms can be used as one measure of the condensate strength.[20] An example of the LDOS vs. energy along the channel is shown in Fig. 3, which exhibits a gradual crossover between regions with and without the condensate. The crossover results from both the evanescent decay of lead states in the gap region, much as would occur at a conventional heterojunction, and the associated non-local self-consistently-calculated exchange interaction that is inherently weakened near the simulation region boundaries.

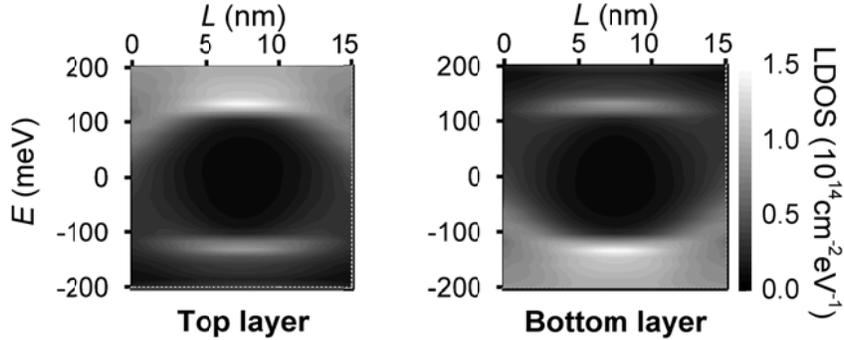

FIG.3. LDOS vs. energy and position within the channel for $\varepsilon_r$ = 2.2, $V_b$ = 0.5 meV, $d$ = 1 nm, $L$ = 15 nm and $n = p \approx 6 \times 10^{12}$ cm$^{-2}$. The lack of LDOS is demonstrated by the dark regions.

**B. Intra-layer and interlayer transmission probabilities**

In the presence of the condensate-induced band gap, the intra-layer transmission through the channel within a single layer with the energy range of the gap will be attenuated substantially, as is perhaps to be expected. Perhaps less intuitively, the majority of the incident electrons are, however, transmitted to the other layer on the same end of the channel in the presence of the condensate. The average transmission probabilities as a function of energy for injections from leads BL and TL with and without the condensate are contrasted in Fig. 4 for $\varepsilon_r$ = 2.2, $d$ = 1 nm, $L$ = 15 nm, $n = p \approx 6 \times 10^{12}$ cm$^{-2}$, and interlayer A-A bare coupling, $H_b(\mathbf{R}_T, \mathbf{R}_B) = V_b = 1$ meV between atoms on the A sublattice of the same bilayer unit cell, and zero otherwise. (Note that different densities of transverse modes in the top and bottom layers, which come from the interlayer electrostatic potential difference, cause the non-equality between *averaged* transmissions over propagating modes from BL to TL and those from TL to BL. However, detailed balance is guaranteed for any given single $k_y$ mode.) In the absence of the condensate, almost all injected current is transmitted to the opposite side of the same layer (Fig. 4(c)) with the maximum interlayer transmission over three orders of magnitude lower (Figs. 4(b) and

(d)). In the presence of the condensate, intra-layer transmission falls to a few percent in the center of the energy gap (Fig. 4(g)), while the majority of the injected current (~75%) tunnels through the interlayer dielectric to the same end of the channel (Fig. 4(f)). The reason for enhanced interlayer transmission probabilities is the condensate-*enhanced* interlayer coupling, composed of the bare coupling as well as the exchange interaction.

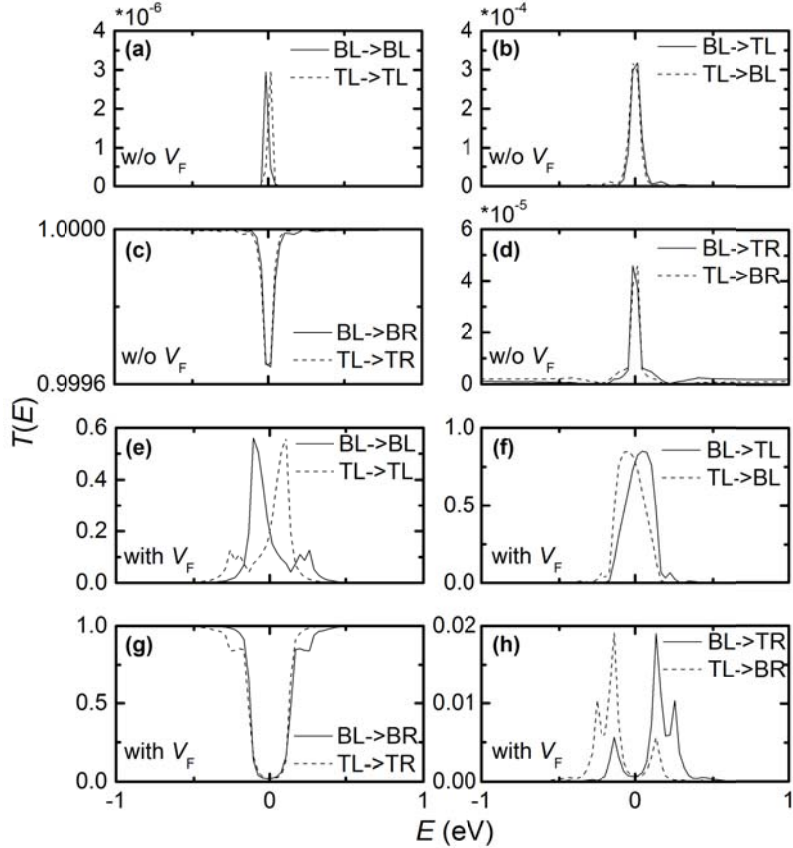

FIG. 4. The value of the transmission coefficients $T(E)$, including that for reflection to the lead of incidence and averaged over all incident modes, for injection from leads BL and TL. Transmission coefficients without the condensate are plotted in (a)-(d), and those with the condensate in (e)-(h). Simulation parameters are dielectric constant $\varepsilon_r = 2.2$, layer separation $d = 1$ nm, channel length $L = 15$ nm, interlayer bare coupling $V_b = 1$ meV and carrier concentrations $n = p \approx 6 \times 10^{12}$ cm$^{-2}$.

### C. Drag-counterflow biasing and exciton flow

As illustrated by the example of in Fig. 5(b) and (d), under the drag-counterflow biasing condition, $V_{BL} = -V_{BR} = V_{al}/2$ and $V_{TL} = V_{TR} = 0$, currents of nearly identical magnitude flow along the upper and lower layers in the channel but in opposite directions. These currents flow despite the lack of any bias on the leads to the top layer, and despite the band gap within the channel in presence of the condensate exhibited in Fig. 3 and the associated low intra-layer equilibrium transmission probabilities of Fig. 4 near the chemical potential. In addition, the interlayer current is relatively limited by comparison despite the large interlayer transmission probabilities of Fig. 4 in the presence of the condensate. To understand this behavior we must delve deeper, literally here in terms of the energy. Figs. 6(d)-(f) and (h) show the current flow subdivided about a cutoff energy $E_{co} = -65$ meV that is both well below the chemical potential(s) and above the nominal lower edge of the condensate-induced band gap at approximately −100 meV (Figs. 2 and 3). The charge current at the energy of the injected carriers is indeed substantially attenuated in the channel, as shown in Figs. 6(d), (e) and (h), although not completely for this short 15 nm channel, and flows primarily between the layers, as shown in Figs. 6(f) and (h). However, an opposing current loop along and between layers within the channel is excited below $E_{co}$—in an entirely elastic process—that essentially vanishes at the ends of the channel, as also shown in Figs. 6(f) and (h). (In that this sub-$E_{co}$ current essentially vanishes at the channel ends and that its peak value almost matches that of the incident and extracted currents, it becomes clear that this sub-$E_{co}$ current is not just the thermal tail of the injected and extracted current distributions.) It is this nominally sub-band-gap current that extends the total current flow all the way through the channel. Unlike for bare coupling alone, the interlayer phase relationship for different states $\beta$ becomes self-consistently tied together through the complex exchange potential that comes through the global interlayer phase relationship present in the interlayer density matrix $\rho(\mathbf{R}_T,\mathbf{R}_B)$ (Eqs. (1)-(3)). It is in this way that injected/extracted current within the condensate-induced band gap can induce current flow below the band gap.

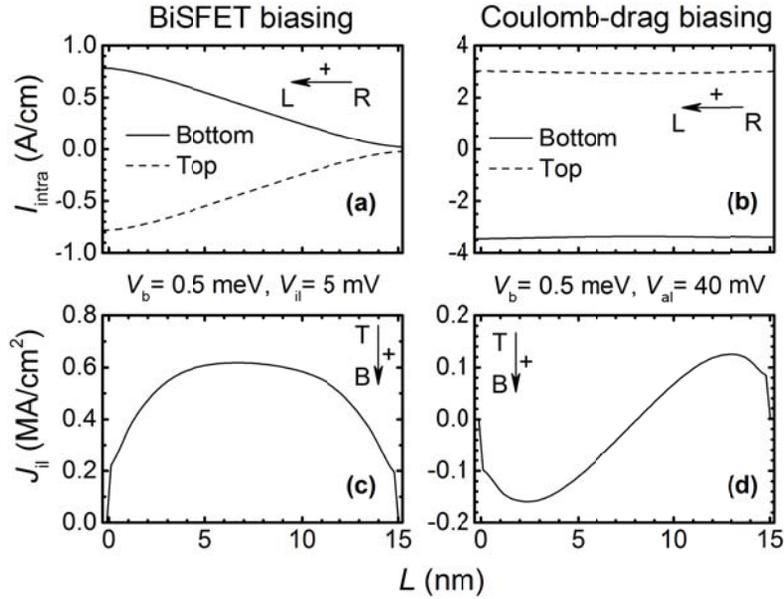

FIG. 5. Illustrative samples for distributions of intra-layer current per width ($I_{intra}$) and interlayer current per area ($J_{il}$). (a) and (c): BiSFET-like biasing with $V_{il} = 5$ mV and $V_b = 0.5$ meV, and (b) and (d): drag-counterflow current biasing with $V_{al} = 40$ mV and $V_b = 0.5$ meV. Other parameters are $\varepsilon_r = 2.2$, $d = 1$ nm, $n = p \approx 6\times10^{12}$ cm$^{-2}$ and $L = 15$ nm. The positive directions for intra- and interlayer currents are as shown in the plots.

While the calculations are performed in terms of propagating electron states only, although within the valence band of the p-type layer, turning to the language of electrons and holes can be helpful in understanding the essential physics. At one end of the channel, the current flow can be interpreted as electrons and holes being injected from (extracted to) the n-type and p-type leads into (from) the end of the channel to create/emit/excite/form (annihilate/absorb/destroy/break-up) coherent excitons and associated exciton flow within the channel from (to) that end. For Coulomb drag-counterflow current biasing, specifically, the interlayer current flow above $E_{co}$ in Figs. 6(f) and (h) is associated with the injection and extraction of electron-hole pairs at the opposite ends of the channel. The oppositely directed interlayer current flow below $E_{co}$ in the same figures is associated with the corresponding creation and annihilation of the coherent excitons. The intra-layer current flow below $E_{co}$ in Figs 6(d),(e) and (h) is associated with the flow of the excitons along the channel, with corresponding opposite *charge* current flows in the two layers, producing the near-perfect Coulomb drag-counterflow current. Moreover, while the electron portion of the exciton pairs flows at the lower edge of the condensate induced band gap within the channel, the hole portion flows at the upper edge of the band gap under the energy sign change required to define holes, which makes the total energy conservation clear even as the energies of the constituent electrons and holes change as they are injected to (extracted from) the channel to create (annihilate) the excitons within the channel. Finally, we note that this process is somewhat analogous to Andreev reflection in which an injected electron and reflected hole at the edge of a conventional superfluid region creates a coherently bound Cooper pair of electrons in an energy conserving manner, but here an injected electron in the n-type layer and a reflected electron (an injected hole) in the p-type layer create a bound electron-hole pair, or an exciton.

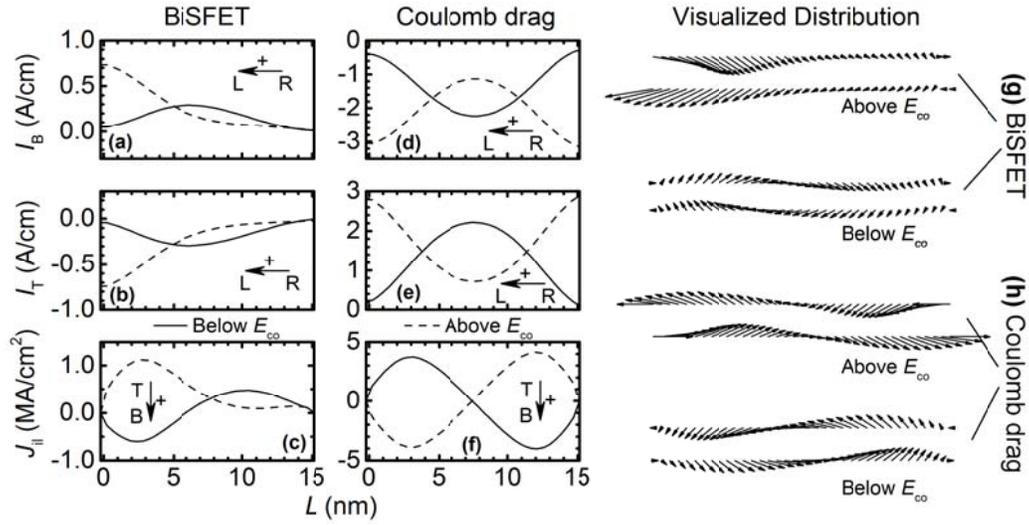

FIG. 6. Bottom-layer ($I_B$), Top-layer ($I_T$) current per width and interlayer ($J_{il}$) current per area distributions for (a)-(c) BiSFET biasing, and (d)~(f) drag-counterflow biasing, using the same system parameters, biasing conditions and rules for signs of current flow as for Fig. 5. These current distributions are also visualized with arrows that originate at the positions they represent for (g) BiSFET-like and (h) drag-counterflow biasing, with vertical and horizontal components proportional to the interlayer and intra-layer currents, respectively. However, with the interlayer and intra-layer current having different units, the shown relative lengths of the horizontal and vertical components are chosen for illustration clarity only. The currents above $E_{co}$ correspond to electron and hole injection from the leads into or extraction to the leads from the channel region, while those below $E_{co}$ correspond to coherent exciton flow within the channel.

Finally, we note that as the bias $V_{al}$ is increased, the near perfect Coulomb drag-counterflow current (Fig. 7(a)) is maintained until ultimately the condensate itself collapses, as shown by the collapse of the band gap in Fig. 7(b). This eventual collapse appears to be with increasing free carrier concentrations above and below the band gap with the voltage-defined splitting of the chemical potentials, which collapses the condensate in much the same way as high temperatures or large charge imbalances would do.[20]

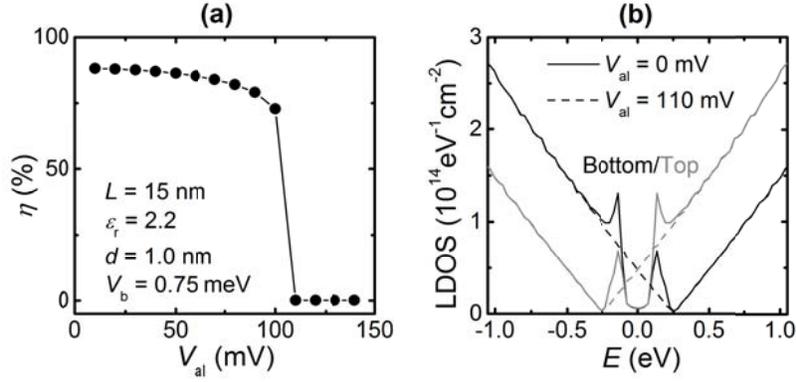

FIG. 7. (a) Collapse of the near-perfect Coulomb drag-counterflow current in the presence of the condensate in a nanoscale channel with parameters provided in the figure. The bottom layer is biased, and thus the drag-efficiency $\eta$ is defined as the top-to-bottom ratio between corresponding current amplitudes averaged across the channel. (b) LDOS in the center of the channel for representative voltages that do and do not support the near-perfect Coulomb drag-counterflow current, as indicated in (a). The collapse of the near-perfect Coulomb drag-counterflow current is accompanied by the collapse of the condensate.

### D. Critical current and voltage within the BiSFET-like biasing scheme

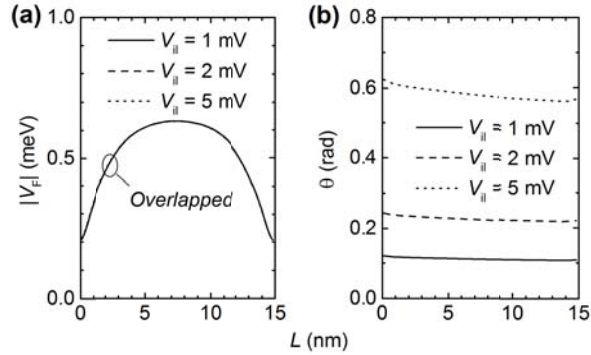

FIG. 8. (a) Amplitude of the Fock exchange interactions (proportional to the pseudospin amplitude) between bare-coupled pairs of top and bottom atoms as a function of position in a 15 nm channel. Other parameters are $V_{\text{diff}} = 0.5$ eV, $V_b = 0.5$ meV, $d = 1.2$ nm and $\varepsilon_r = 2.2$. (b) Total pseudospin phase between the same pairs of atoms as a function of position with the same system parameters. Apparently, the pseudospin amplitude is position-dependent yet voltage-independent, while the pseudospin phase is voltage-dependent yet largely position-independent.

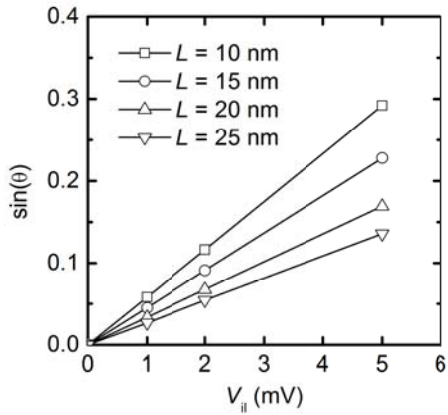

FIG. 9. Linear $\sin(\theta)$ vs. $V_{\text{il}}$ relation for BiSFET biasing conditions, with $\varepsilon_r = 2.2$, $d = 1$ nm, $V_b = 0.75$ meV, and $n = p \approx 6 \times 10^{12}$ cm$^{-2}$, for varying channel lengths $L$.

For the BiSFET-like biasing condition, $V_{\text{TL}} = -V_{\text{BL}} = V_{\text{il}}/2$, $V_{\text{TR}} = V_{\text{BR}} = 0$, we find that condensate formation can greatly increase the interlayer conductance, reaching ~75% of the Landauer-Büttiker ballistic limit for the leads, consistent with expectations based on the interlayer transmission probabilities. Remember that the equilibrium interlayer transmission probabilities are orders of magnitude lower (Fig. 4) without the condensate. (The BiSFET nominally has contacts to only one end, while here a negligible current flows to the grounded contacts for the "BiSFET-like" biasing, as seen in Fig. 5(a), so that the systems are the same in terms of this aspect of current flow.) However, this high, at least, DC conductance can only be maintained up to a critical current $I_c$, again as determined by self-consistent effects of the current flow on the entire interlayer density matrix $\rho(\mathbf{R}_T,\mathbf{R}_B)$. The interlayer density matrix $\rho(\mathbf{R}_T,\mathbf{R}_B)$ may be defined as a collective "pseudospin" $\rho(\mathbf{R}_T,\mathbf{R}_B)$ with pseudospin magnitude $|\rho(\mathbf{R}_T,\mathbf{R}_B)|$ and phase $\theta(\mathbf{R}_T,\mathbf{R}_B) \equiv \arg[\rho(\mathbf{R}_T,\mathbf{R}_B)]$, which is analogous to a collective real spin state—magnetic moment—with strength and orientation. (This terminology underlies the BiSFET and BiSJT names.) As illustrated by Fig. 8, under BiSFET-like biasing, the pseudospin magnitude exhibits a position dependence consistent with the localization of the condensate, but is essentially independent of the interlayer bias, while the pseudospin phase is weakly dependent of position but varies with $V_{il}$, $\theta(\mathbf{R}_B,\mathbf{R}_T) \approx \theta$.[13] With $\theta$ evaluated at mid-channel for specificity, $\sin(\theta)$ is proportional to $V_{il}$ as shown in Fig. 9. The inter-atom current flow between layers of Eq. (10) can be written as

$$I(\mathbf{R}_T,\mathbf{R}_B) \cong I_{max}(\mathbf{R}_T,\mathbf{R}_B)\sin(\theta),\tag{11a}$$

where

$$I_{max}(\mathbf{R}_T,\mathbf{R}_B) = (-2e/h)H_b(\mathbf{R}_T,\mathbf{R}_B)|\rho(\mathbf{R}_T,\mathbf{R}_B)|.\tag{11b}$$

Similarly, the total interlayer current $I$ can be written as

$$I \cong I_c \sin(\theta),\tag{12a}$$

where, in general,

$$I_c = \sum_{\mathbf{R}_T}\sum_{\mathbf{R}_B} I_{max}(\mathbf{R}_T,\mathbf{R}_B),\tag{12b}$$

although, as previously noted, in this work we only include bare coupling, and thus total inter-atomic-site current flow, between top- and bottom-layer A atoms in the same unit cell. With $\sin(\theta)$ having a maximum magnitude of unity when $|\theta| = \pi/2$, the maximum or "critical" current magnitude is reached, as indicated by the "max" and "c" subscripts in Eqs. (11) and (12). In practice, as $|\theta|$ approaches $\pi/2$ in our simulations, the rate of convergence slows asymptotically.[17] Therefore, we use the observed linear dependence of $\sin(\theta)$ on $V_{il}$ to extract the critical currents $I_c$ and corresponding values of associated "critical voltage" $V_c$ from somewhat smaller interlayer currents and voltages.[13] (Such linearity is observed also at $|\theta|$ close to $\pi/2$ in Ref. 17.) Thus-extracted $V_c$ are shown in Fig. 10 as a function of the bare interlayer coupling, and for $L = 15$ and $20$ nm.

It was not possible to converge any solution in our steady-state calculations beyond the critical voltage. Instead, the solutions are not numerically stable in a conventional sense. Instead, the pseudospin magnitude remains essentially constant with iteration while the phase smoothly rotates through all angles periodically, the larger $V_{il}$ beyond $V_c$, the faster the rotation with iteration.[14,17] The nearly constant pseudospin magnitude suggests that the condensate does not collapse beyond $V_c$, in contrast to cases for the collapse of the Coulomb drag-counterflow current (Fig. 7). The rotation with iteration suggests instead time-dependent current oscillations analogous to those in superconducting Josephson junctions, which have voltage-dependent DC currents below a critical value followed by collapse of the DC current and an onset of an AC current of fixed amplitude after the critical current is reached.[11,25] Similarly, we expect an oscillation rate $f$ of $f/(V_{il} - V_c) \approx 2q/h \approx 0.5$ THz/mV, which would be well into the THz regime for the simulated systems here. Consistent with the "pseudospin" terminology, this transition from steady-state to oscillatory behavior under increasing drive voltage also is analogous to that expected for in-plane easy axis nanomagnets (magnets with in-plane magnetic anisotropy) driven by spin transfer torque (STT) via interlayer charge current flow through perpendicular easy axis nanomagnets.[26] However, to directly model the behavior beyond $V_c$ would require time-dependent simulations beyond our current capabilities.[22]

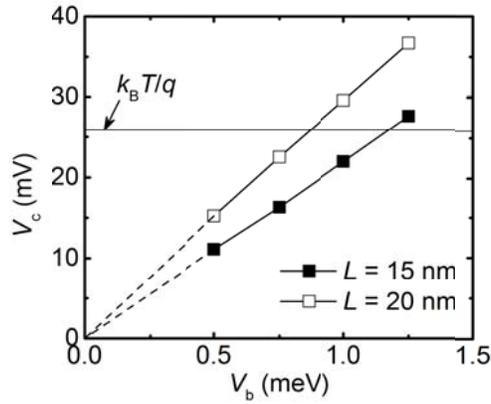

FIG. 10. Linear dependence of critical voltage $V_c$ on interlayer bare coupling energy $V_b$. System parameters are $n = p \approx 6 \times 10^{12}$ cm$^{-2}$, $\varepsilon_r$ = 2.2, and $d$ = 1 nm.

The critical currents and, as illustrated in Fig. 10, the associated critical voltages $V_c$ also are linearly dependent on the bare interlayer hopping energy $V_b$, consistent with Eq. (11b). Physically, this dependence can be understood in the following way. Above $E_{co}$ most of the current flow is between layers, but again a current loop is excited in the channel below $E_{co}$ (Figs. 6(a)-(c) and (g)) that is associated with exciton flow. With incident (outgoing) electron and holes creating (annihilating) excitons at one end of the channel but not extracted (injected) at the other end, a steady-state exciton population can only be maintained through bare-coupling-assisted recombination (generation) within the channel. Therefore, the greater the bare coupling, the greater current of injected (outgoing) electrons and holes that can be maintained in the opposite layers. Conversely, with excitons injected at one end of the channel and extracted at the other under drag-counterflow biasing, much larger current injection and extraction can be supported with the same or lower $V_b$ as already seen (Fig. 5(b) vs. Fig. 5(a), e.g.).

Critical for application to the proposed BiSFET and BiSJT, the calculated critical voltages can be scaled below the thermal voltage $k_BT/q$ (where $k_B$ is Boltzmann's constant), which is approximately 26 mV in these 300 K simulations, via the linear dependence on the bare coupling, as in the examples shown in Fig. 10. The basis for these sub-$k_BT/q$ critical voltages is the collective effect, represented here through the Fock exchange interaction that couples many electron states $\beta$ together.

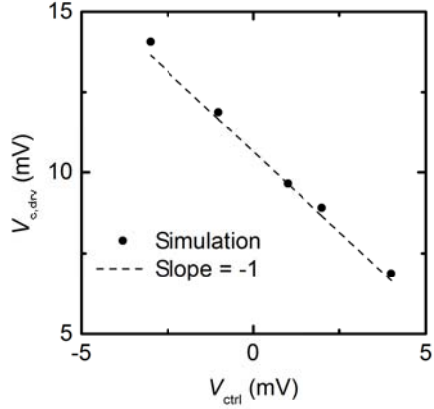

FIG. 11. Effective interlayer critical voltage on the "drive" (drv) end of the channel at which the critical interlayer current is reached as a function of the interlayer voltage applied to the "control" (ctrl) end of the channel. The dashed line with slope of −1 corresponds to perfect critical voltage conservation such that sum of the two voltages at which the critical current is reached would be fixed. System parameter are $\varepsilon_r = 2.2$, $V_b = 0.5$ meV, $L = 15$ nm, $d = 1.0$ nm and $n = p \approx 6\times10^{12}$ cm$^{-2}$.

### E. Mixed BiSJT biasing

While biasing both ends of the channel can allow for more intra-layer current flow, it does not increase the ability to drive interlayer current, at least not a net current. Instead, when we mix the BiSFET-like and drag-counterflow biasing schemes to obtain the BiSJT regime, $V_{TL} = -V_{BL} = V_{ctrl}/2$, $V_{TR} = -V_{BR} = V_{drv}/2$, we observed behavior that is, indeed, the basis for the proposed BiSJT[15] from which we have borrowed the "control" (ctrl) and "drive" (drv) nomenclature. Specifically, we find that the total interlayer critical current and associated total voltage applied to both channel ends are essentially conserved, as shown in Fig. 11. Equivalently, application of a sub-critical voltage on one end with an associated sub-critical interlayer current can change the *effective* critical current and voltage seen at the other end. In the 1D-effective-mass based simulations of Ref. 24, it was found that bare interlayer coupling was necessary to allow less-than-perfectly matched counterflow intra-layer currents, which is a different perspective on the same basic conclusion.

### F. Channel scaling, system parameters, bulk critical temperature and normalized wavelength

Interlayer critical current, voltage and the conductance have been calculated with various system parameters such as channel length $L$, interlayer bare coupling energy $V_b$, interlayer separation $d$, effective dielectric constant $\varepsilon_r$, and carrier concentrations $n$ and $p$. All the simulations are based on the BiSFET-like biasing scheme, $V_{TL} = -V_{BL} = V_{il}/2$ and $V_{TR} = V_{BR} = 0$, without losing generality given the critical current and voltage conservations. For reference we also use the bulk value of the critical temperature at which the condensate collapses, $T_c$, which is a measure of the latter's strength, as a unifying parameter. (As well as being calculated directly) $T_c$ can be estimated from the 0K induced bulk band gaps $E_{g0}$ as $T_c \approx E_{g0}/(4k_B)$ essentially independent of dielectric environment, interlayer separation or carrier concentration.[20] We also use $\lambda$, the Fermi wavelength normalized to the transition length required to form the condensate in the channel, to characterize the abruptness of the channel in a way that is relevant to quantum mechanical reflection probabilities. For this purpose, the transition length is defined as the distance across which the LDOS at the chemical potential drops from 90% to 10% of its value in the absence of the condensate. These two parameters are interdependent in that the bulk condensate strength affects how abruptly the condensate can form with position in the channel, as will be seen below.

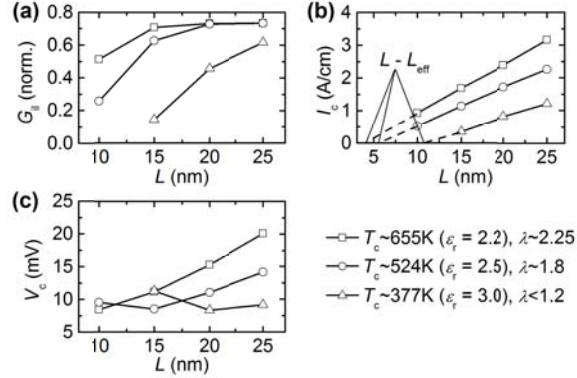

FIG. 12. Interlayer conductance $G_{il}$ normalized to Laudauer-Büttiker limit for the leads at the chemical potential (0 eV), interlayer current density $I_c$ (per unit width) and critical voltage $V_c$ with varying effective dielectric constant $\varepsilon_r$ = 2.2, 2.5 and 3.0 and corresponding bulk critical temperatures $T_c \approx$ 655K, 524K and 377 K, respectively. (For $\varepsilon_r$ = 3.0 ($T_c \approx$ 377 K) the condensate is only partially formed within the 25 nm channel, and therefore the transition length cannot be determined by the definition in the text. Therefore, $\lambda$ is taken to be simply less than the Fermi wavelength divided by half the maximum channel length (12.5 nm), which is 1.2.) The other parameters are $V_b$ = 0.5 meV, $d$ = 1 nm and $n = p \approx 6 \times 10^{12}$ cm$^{-2}$.

We first consider what is a voltage-independent interlayer conductance $G_{il}$ below the critical current $I_c$ and critical voltage $V_c$ as a function of channel length for differing effective dielectric constants, $\varepsilon_r$ = 2.2, 2.5 and 3.0, as in Fig. 12 with the corresponding $T_c$ and $\lambda$ listed in the collective figure inset. Other parameters are $d$ = 1 nm, $V_b$ = 0.5 meV and $n = p \approx 6 \times 10^{12}$ cm$^{-2}$. As seen in Fig. 12(a), with increasing channel length, $G_{il}$ increases and saturates to about 75% of the Landauer-Büttiker limit $G_{LB}$ for the leads, to which it is normalized here. ($G_{LB} \cong (V_{diff}/t_0)(1.15$ kS/cm), where $t_0$ = 2.7 eV is the magnitude of the nearest-neighbor intra-layer bare coupling potential in the tight-binding model).

The rate at which $G_{il}$ saturates increases with the strength of the condensate as measure by $T_c$. Consistent with bulk estimates,[20,21] Fig. 12(b) exhibits a linear dependence of interlayer critical current $I_c$ (actually a current density per unit width for this figure and those to follow) on channel length $L$, suggesting a constant critical current per unit area if a shorter *effective* channel length $L_{eff}$—the actual $L$ minus its $L$ axis intercept for each linear $I_c$ vs. $L$ curve—is considered to allow for the non-abrupt condensate formation illustrated in Fig. 3.[20] $L_{eff}$ increases with $T_c$. The difference between $L$ and $L_{eff}$ lies in the non-local nature of the exchange interaction. With a half-width (radius) $\mathbf{R} = |\mathbf{R}_T - \mathbf{R}_B|$ of roughly 2 nm at half maximum of bulk density matrix elements $\rho(\mathbf{R}_T, \mathbf{R}_B)$ for this graphene system, defined by both the Coulomb interaction and coherence length of the interlayer density matrix,[20] a gradual build-up of the condensate within the channel is inherent. Moreover, in determining the net strength of the exchange interaction in, e.g., the top layer at some point $\mathbf{R}_T$ within the transition region, a reduced amplitude in $\rho(\mathbf{R}_T, \mathbf{R}_B)$ for coupling to points $\mathbf{R}_B$ closer to the channel edge must be compensated for by an increased amplitude in $\rho(\mathbf{R}_T, \mathbf{R}_B)$ for coupling to points closer to the channel center. The latter becomes increasingly difficult to obtain as the critical temperature $T_c$ approaches the 300 K simulation temperature. Thus, $L - L_{eff}$ increases with decreases in $T_c - T$. (The relationship of the non-local nature of the exchange interaction to $L - L_{eff}$ also illustrates the need for explicit modeling of the non-local interaction to judge the potential for nano-scale condensates.) In addition, the effective critical current density, $J_{c,avg} = I_c/L_{eff}$, increases with $T_c$ even allowing for the increases in $L_{eff}$ with $T_c$, consistent with the dependence of $I_{max}(\mathbf{R}_T,\mathbf{R}_B)$ on $|\rho(\mathbf{R}_T,\mathbf{R}_B)|$ (Eq. 12(b)).

At this point, we note that for computational expediency we generally have considered very strong condensates, with the lowest $T_c$ of 377 K still not producing $G_{il}$ saturation with a 25 nm channel length in this modeled system at 300 K. However, we also note that saturation is not necessary to achieve greatly enhanced $G_{il}$ still on the scale of the Landauer-Büttiker limit (Fig. 12(a)) and orders of magnitude greater than the condensate-free interlayer conductance expected based on the transmission probabilities of Fig. 4. Moreover, saturation may not even be ideal from one perspective. The critical voltage, $V_c = I_c/G_{il}$, decreases with $L$ initially due to the faster increase in $G_{il}$ than in $I_c$, and increases with $L$ afterwards because of the saturation of $G_{il}$ and continuous increase of $I_c$. Thus, the point at which $V_c$—an important parameter for BiSFET and BiSJT circuit applications—is the least sensitive to variations in channel length is reached prior to saturation.

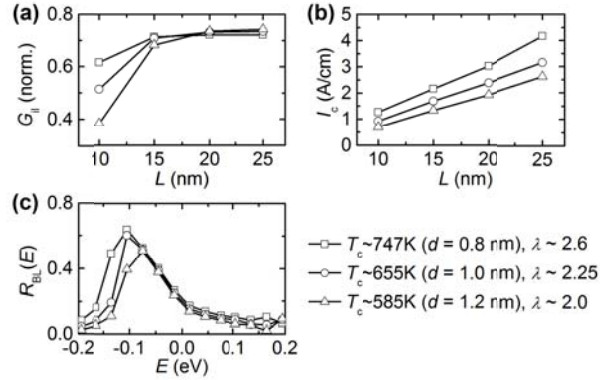

FIG. 13. Interlayer conductance $G_{il}$ normalized to Laudauer-Büttiker limit for the leads, interlayer critical current density $I_c$ per unit width, and (non-equilibrium) reflection probabilities, for incident electrons in the bottom-left lead BL, $R_{BL}$ for varying interlayer separation $d$. The other parameters are $\varepsilon_r = 2.2$, $V_b = 0.5$ meV and $n = p \approx 6 \times 10^{12}$ cm$^{-2}$. The interlayer separations $d$ are 0.8, 1.0 and 1.2 nm, with associated bulk critical temperatures of $T_c \approx 747$ K, 655 K and 585 K, respectively.

For the simulation results of Fig. 13, differing interlayer separations, $d = 0.8$, 1.0 and 1.2 nm, are considered with $\varepsilon_r = 2.2$, $V_b = 0.5$ meV, and $n = p \approx 6 \times 10^{12}$ cm$^{-2}$. The basic trends with respect to $T_c$ are much the same for both $G_{il}$ saturation and $I_c$ with $L$ as seen in Fig. 12. However, the spread in $T_c$ is less as expected given its generally weaker dependence on $d$ than to $\varepsilon_r$,[20] despite the slightly larger percentile variation in $d$ considered here, and thus in the results as well. In addition, we see that that the saturated value of $G_{il}$ can vary, although the overall effect is not large in this case. While the smaller the layer separation the more quickly $G_{il}$ saturates as expected, the saturation is slightly larger for larger $d$ (Fig. 13(a)). This increase in saturated $G_{il}$ is associated with less abrupt change in the potential upon entering the channel—albeit the interlayer exchange mediated hopping energy—indicated by smaller $\lambda$. The averaged reflection coefficients as a function of energy (Fig. 13(c)), in the vicinity of the chemical potential (0 eV) where the incident and outgoing currents flow, exhibit this effect even more strongly. (The reflection probability decreases above the chemical potential as the energy of incidence approaches the condensate band edge allowing more transmission, and initially increases below because of a lack of available final states in the opposite layer.) The overall effect, however, is not large.

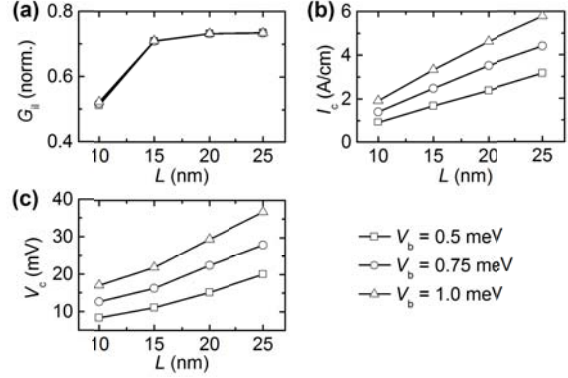

FIG. 14. Interlayer conductance $G_{il}$ normalized to Laudauer-Büttiker limit for the leads, interlayer critical current density $I_c$ per unit width, and associated critical voltage $V_c$ for varying interlayer hopping energies $V_b = 0.5, 0.75$ and $1.0$ meV. The other parameters are $\varepsilon_r = 2.2$, $d = 1$ nm and $n = p \approx 6\times10^{12}$ cm$^{-2}$. $T_c \sim 655$K and $\lambda \sim 2.25$ for all three cases.

For the results of Fig. 14, different interlayer bare coupling strengths were considered, $V_b = 0.5, 0.75$ and $1$ meV, with $\varepsilon_r = 2.2$, $d = 1$ nm, and $n = p \approx 6\times10^{12}$ cm$^{-2}$. The strength of the condensate and associated $T_c$ are not significantly affected by $V_b$ at this scale, and, thus, the interlayer conductance $G_{il}$ also remains essentially unaffected (Fig. 14(a)). However, the critical current (Fig. 14(b) and associated critical voltage (Fig. 14(c)) increase with increasing $V_b$, consistent with Eq. 11(b). For strong bare coupling, past bulk work has shown that the strength and even the detailed shape of the interlayer density matrix $\rho(\mathbf{R}_T,\mathbf{R}_B)$ can be affected.[21] However, such strong interlayer bare coupling also would likely produce $V_c$ substantially greater that $k_BT/q$.

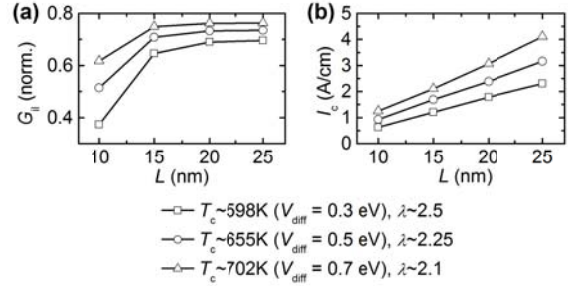

FIG. 15. Interlayer conductance $G_{il}$ normalized to Laudauer-Büttiker limit for the leads, and interlayer critical current density, $I_c$ per unit width, with varying carrier concentrations $n = p \approx 3.8\times10^{12}$ cm$^{-2}$, $6.0\times10^{12}$ cm$^{-2}$ and $8.6\times10^{12}$ cm$^{-2}$ ($V_{diff} = 0.3, 0.5$ and $0.7$ eV respectively) and corresponding bulk critical temperatures $T_c \approx 598$ K, $655$ K and $702$ K, respectively.

Finally, we consider differing nominal carrier concentrations, $n = p \approx 3.8\times10^{12}$, $6.0\times10^{12}$, and $8.6\times10^{12}$ cm$^{-2}$ ($V_{diff} = 0.3, 0.5$ and $0.7$ eV, respectively) in Fig. 15, with $\varepsilon_r = 2.2$, $V_b = 0.5$ meV, and $d = 1.0$ nm. These parameter sets exhibit the largest difference in saturated $G_{il}$ as normalized to the Landauer-Büttiker limit for the corresponding carrier concentration. Here, the decrease in the Fermi wavelength relative to the valley center accompanying the increase in carrier concentration exceeds the decrease in transition length caused by stronger condensates, producing a smaller wavelength-normalized potential change (smaller $\lambda$). Moreover, this smoother potential change is associated with a stronger condensate. The combination produces a significant increase in the saturated normalized $G_{il}$ (Fig. 15(a)). Of course, the Landauer-Büttiker limit for $G_{il}$ also increases with carrier concentration. By comparison, it is likely that resulting smoother potential change and weaker condensate in the cases for Figs. 12 and 13 partially compensate each other producing smaller if any changes in the saturated normalized $G_{il}$. In addition, the stronger condensate leads to a larger critical current per channel length (Fig. 15(b)) or even effective channel length.

## IV. CONCLUSION

We have developed an atomistic tight-binding NEGF quantum transport method (except that we need not solve for the full Green's function matrix in these ballistic simulation) for modeling non-equilibrium transport through nanoscale exciton condensates realized via the non-local many-body Fock exchange interaction. Specifically, we have modeled the graphene-dielectric-graphene system here. However, the essential physics should be similar in TMD-based material systems also under consideration or other material systems. In addition, the method itself should also be extendable to these latter systems, although with a greater computational burden. We have exhibited the possibility for condensate formation within nanoscale regions, and its dependence on the critical temperature of the bulk condensate. We have exhibited essentially transport effects that serve as the basis for beyond-CMOS device proposals including: interlayer conductances approaching the Landauer-Büttiker limits imposed by the leads but limited by critical currents (beyond which collapse of DC conductance and onset of THz AC conductance are expected, which is at least consistent with the oscillation in the pseudospin phase with iteration seen here), with the latter reached at sub-$k_BT/q$ voltages (BiSFET); critical current conservation for currents injected into two—or likely more—regions of the condensate (BiSJT); and near-perfect Coulomb drag-counterflow current between layers. We also have exhibited the underlying transport physics including the process by which incident (outgoing) electron-hole pairs in opposite layers excite (absorb) coherent excitons, in a manner somewhat analogous to Andreev reflection at the edge of conventional superfluids. The work is not intended to address the likelihood of achieving such condensates, which requires much greater attention to the details of screening and ultimately must be resolved experimentally. It is intended to motivate such work through exhibiting the novel transports effect that could thus be achieved, and perhaps to help with the interpretation of such experimental results.